\documentclass[eps,preprint,12pt]{revtex4}

\usepackage{amsmath}
\usepackage{amssymb}
\usepackage{amsfonts}
\usepackage{graphicx}
\usepackage{dcolumn}
\usepackage{bm}
\usepackage{hyperref}
\hypersetup{colorlinks=false}

\begin{document}

\title{Ho\v{r}ava-Lifshitz Gravity Effects on Casimir Energy in Weak Field Approximation and Infrared Regime}

\author{C. R. Muniz}

\address{Grupo de F\'isica Te\'orica (GFT), Universidade Estadual do Cear\'a, FECLI, Iguatu, Cear\'a, Brazil.}

\author{V. B. Bezerra}

\address{Departamento de F\'{i}sica, Universidade Federal da Para\'{i}ba, Caixa Postal 5008, CEP 58051-970, Jo\~{a}o Pessoa, PB, Brazil}

\author{M. S. Cunha}

\address{Grupo de F\'isica Te\'orica. Universidade Estadual do Cear\'a, Av. Paranjana 1700, CEP 60740-000, Fortaleza, Cear\'a, Brazil.}

\begin{abstract}
We calculate the renormalized vacuum energy of a massless scalar field confined between two nearby parallel plates formed by ideal uncharged conductors, placed tangentially to the surface of a sphere with mass $M$ and radius $R$. This study will take into account the static and spherically symmetric solution of Ho\v{r}ava-Lifshitz gravity found by Kehagias-Sfetsos (KS), in both weak field and infrared (IR) limits. A slight amplification of the Casimir force between the conducting plates is found. Thermal corrections to the Casimir energy are analyzed. Based on current Casimir effect measurements, a constraint on the $\omega$ parameter of KS metric is also obtained.

\vspace{0.75cm}
\end{abstract}


\maketitle

\section{Introduction}

 The Ho\v{r}ava-Lifshitz gravity offers the possibility to construct a viable theory of quantum gravity due to the fact that this theory is power-counting renormalizable \cite{horava1,horava2,horava3}. In order to get this renormalizability, the standard 4-dimensional invariance by diffeomorphisms and the local Lorentz symmetry are abandoned and a different kind of scaling at very short distances, i.e., in the ultraviolet (UV) regime, is adopted, with those symmetries emerging accidentally in an infrared (IR) regime \cite{visser}. In the UV scale, spacetime is therefore broken, occurring an anisotropic scaling of time as $t\rightarrow b^zt$ and space as $x^{i}\rightarrow bx^{i}$, in which $z$ is a critical exponent that tends to unity at large distances. This restores the validity of General Relativity in IR scale, or at least Lorentz violations in this scale occurs below current experimental constraints. In three-dimensional space, the renormalizability of the theory exists for $z=3$, and, in general, it occurs for $z=d$ in a $d$-dimensional space.

Ho\v{r}ava-Lifshitz gravity has smaller strength at high energy scale than General Relativity, due to the presence of $z$ order spatial derivatives in the action, and at the same time it is ghost-free by demanding only first order temporal derivatives. Such aspects generate a bouncing cosmology and an accelerating universe, i.e, dark energy \cite{mukohyama}. Furthermore, the cosmological constant present in Ho\v{r}ava-Lifshitz theory is very large and negative, and can solve the problem of huge discrepancy between the observed value of the cosmological constant and standard predictions from Quantum Field Theory, that indicate a large and positive value for the global vacuum energy of matter \cite{corrado}. Another model of universe, the G\"odel-type one, was also investigated in the context of Ho\v{r}ava-Lifshitz gravity, with the conclusion that the non-causal structure of the G\"odel model is not allowed in that theory \cite{fonseca}.

On the other hand, we hope that the local quantum vacuum of the fields can be also affected by the fundamental separation between time and space that exists at the core of the Ho\v{r}ava-Lifishitz theory. In a recent paper \cite{ferrari}, modifications in Casimir energy of the parallel plates configuration were analyzed in the context of that theory. The considered vacuum was that of a massless scalar field in flat spacetime, and the Ho\v{r}ava proposal was implemented in such manner that the kinetic term of the Lagrangian acquired higher order spatial derivatives, in both perturbative and non-perturbative approaches. This brought unsatisfactory results from experimental point of view, namely, change of signal of the Casimir force in the non-perturbative case (for $z=3$) and persistence of a residual correction even at IR regime (for $z=1$) in the perturbative one.

The Casimir effect is a physical phenomenon which corresponds to an attractive force arising between two parallel, uncharged metallic plates in vacuum \cite{casimir1,bordag}. It is a purely quantum effect resulting from the modification of the zero-point (vacuum) oscillations of the electromagnetic field by the material boundaries if compared with the empty Minkowski space. The Casimir effect also can be present in empty spaces with non-Euclidean topology \cite{DeWitt,Ford}. In such spaces there are no material boundaries, but identification conditions imposed on the quantum fields play the same role as the one introduced by material boundaries. Therefore, it is necessary to take into account properly the Casimir effect in order to have a precise description of the physics of a given system, as for example, the evolution of the universe, or the physics of black holes in different theories of gravity.

In the present paper, we will follow a route in which we will take into account a gravitational background correspondent to a static and spherically symmetric particular solution of Ho\v{r}ava-Lifshitz gravity, namely, the Kehagias-Sfetsos (KS) one \cite{kehagias}. This solution was found without demanding the projectability condition, which consists in the lapse function to be independent of space coordinates, assumed in the original proposal of Ho\v{r}ava's theory \cite{horava1,horava2,horava3}. Due to its simplicity, it is possible to consider the KS solution useful to understand several phenomenological implications of Ho\v{r}ava-Lifshitz gravity. This solution is in agreement with the classical tests of General Relativity \cite{harko}. Other exact spherically symmetric solutions for the Ho\v{r}ava-Lifshitz gravity can be found in \cite{pope,park,kiritsis}.

We will consider the IR limit, such that the gravitational background field used in this paper differs slightly from the Schwarzschild solution, i.e., it will be assumed as a perturbation of the latter in first order of $1/\omega$, where $\omega$ is a parameter that regulates the IR behavior. In the limit in which $\omega\gg1$, we will take $z\rightarrow1$, and the massless scalar field must obey the local Lorentz symmetry. We will calculate its regularized vacuum energy following the procedure used in \cite{sorge}, which obtained the corrections to the flat Casimir energy of a massless scalar field in the configuration of nearly parallel, conducting and discharged plates, which were placed close to the surface of a static spherical body with mass $M$ and radius $R$. The separation between them, $L$, is much small than the radius $R$. We also will calculate the thermal corrections to the Casimir energy, supposing that the plates are at a thermal bath, using the renormalized Helmholtz free-energy, as well as the internal energy in both low and high temperature limits. With these calculations we can verify the dependence of the Casimir energy on the background gravitational field, and how the IR limit influences the final result.

The importance of taking into account the possible Ho\v{r}ava-Lifshitz gravity modifications on the Casimir effect, besides the fact that there are very few studies devoted to the theme, is justified due to the possibility of detection and measurement of that phenomenon in laboratories on Earth. Including effects of temperature can also be relevant, since that, actually, such thermal effects on the interaction between the conducting parallel plates seem to be important for separations greater than $3\mu m$ \cite{khana}.

The paper is organized as follows. In section 2, we present the Kehagias-Sfetsos solution and do the necessary coordinate transformations and approximations in order to adapt it to our problem. In section 3, we calculate the regularized vacuum energy (Casimir energy). In section 4 we calculate the thermal corrections and, finally, in section 5 we discuss the results.

\section{The Static Black Hole Solution in Ho\v{r}ava-Lifshitz Gravity}

We will start with the KS black hole solution, given by \cite{kehagias}
\begin{equation}\label{1}
ds^2=f_{KS}(\rho)dt^2-\frac{d\rho^2}{f_{KS}(\rho)}-\rho^2d\Omega^2,
\end{equation}
where
\begin{equation}\label{2}
f_{KS}(\rho)=1+\omega \rho^2\left(1-\sqrt{1+\frac{4M}{\omega \rho^3}}\right),
\end{equation}
where $\rho$ is a radial coordinate and $\omega$ is a free parameter of the Ho\v{r}ava-Lifshitz theory. Note that when $\omega\rightarrow\infty$ (IR limit), the metric coefficients of (\ref{1}) tend to those of Schwarzschild solution, i.e., $f_{KS}(\rho)\rightarrow f_{Sch}(\rho)=1-2M/\rho$, and that metric turns to the vacuum solution generated by a spherical source, which is consistent with the validity of the General Relativity in IR scale.

We will make the analysis of the Casimir effect taking into account a static background gravitational field, described by the KS solution, whose metric contains a singularity at $\rho=0$. It is easy to see from (\ref{2}) that it has two event horizons, one at $\rho_{out}=M(1+\sqrt{1-1/2\omega M^2})$ and another at $\rho_{inn}=M(1-\sqrt{1-1/2\omega M^2})$. We suppose that the radius $R$ of the spherical body, in whose surface the Casimir apparatus is placed, is much larger than the radii of the event horizons. In other words, we will work in a weak field approximation.

We already have emphasized that our approach will occur close to the IR limit ($z\approx1$, $\omega\gg1$). Thus, we will make an expansion in the metric coefficient given in (\ref{2}) so that
\begin{equation}\label{3}
f_{KS}(\rho)\approx1-\frac{2M}{\rho}+\frac{2M^2}{\omega\rho^4}.
\end{equation}
The last term in (\ref{3}) is what effectively incorporates the properties of Ho\v{r}ava-Lifshitz gravity. We can note that the attractive Newtonian potential is corrected by the term of repulsive character $M^2/\omega\rho^4$.

For our proposes, it is convenient to use of the isotropic coordinates. Thus, it is necessary to find a transformation in the radial coordinate, $\rho=\rho(r)$, such that the KS solution becomes
\begin{equation}  \label{4}
ds^2=g(r)dt^2-h(r)(dr^2+r^2d\Omega^2),
\end{equation}
where $g(r)= f_{KS}(\rho(r))$ and $h(r)=\rho'(r)^2/f_{KS}(\rho(r))=\rho^2/r^2$. The comma (') denotes the derivative with respect to $r$. We have thus a separable ordinary differential equation whose solution is
\begin{equation}\label{5}
\log{(r)}=\int \frac{d\rho}{\rho\sqrt{1-\frac{2M}{\rho}+\frac{2M^2}{\omega\rho^4}}}.
\end{equation}
This integral is of elliptic type, and has no an exact primitive function. We can overcome this difficulty by expanding the integrand up to $\mathcal{O}(M/\rho)$ and $\mathcal{O}(M^4/\psi_0\rho^4)$, where $\psi_0=M^2\omega$ is an adimensional parameter. With this expansion, we are neglecting the higher order effects of Schwarzschild metric and considering the lower order effects of KS solution (\ref{1}). Thus equation (\ref{5}) becomes
\begin{equation}\label{6}
\frac{r}{\rho}\thickapprox\exp{\left(-\frac{M}{\rho}+\frac{M^2}{4\omega\rho^4}\right)}.
\end{equation}
Expanding once more the argument of the exponential in (\ref{6}), we find, finally
\begin{equation}\label{7}
r\thickapprox\rho\left(1-\frac{M}{\rho}+\frac{M^2}{4\omega\rho^4}\right),
\end{equation}
and thus metric (\ref{4}), for $r=R$, can be written as \cite{iorio1}
\begin{eqnarray}\label{8}
ds^2=\left(1-\frac{2M}{R}+\frac{2M^2}{\omega R^4}\right)dt^2-\left(1+\frac{2M}{R}-\frac{M^2}{2\omega R^4}\right)(dr^2+r^2d\Omega^2),
\end{eqnarray}
where we have introduced the isotropic coordinates in order to facilitate the introduction of the parallel plates geometry, in such way that $dr^2+r^2d\Omega^2=dx^2+dy^2+dz^2$.

\section{The regularized vacuum energy in Ho\v{r}ava-Lifshitz gravity}

In this section we will calculate the regularized vacuum energy of the massless scalar field in the weak field approximation and IR regime. To do this, we will write metric (\ref{8}) in the following form

\begin{equation}\label{09}
ds^2=(1+2b\Phi_0')dt^2-(1-2\Phi_0')(dx^2+dy^2+dz^2),
\end{equation}
where
\begin{eqnarray}\label{10}
b&=&1-\frac{3M}{4\omega R^3}\\
\Phi_0'&=&-\frac{M}{R}+\frac{M^2}{4\omega R^4}
\end{eqnarray}

The Klein-Gordon equation for the massless scalar field in the spacetime given by equation (\ref{09}) is written as
\begin{equation}\label{12} \frac{1}{\sqrt{-g}}\partial_{\mu}(\sqrt{-g}g^{\mu\nu}\partial_{\nu})\Psi=[1-2(b+1)\Phi_0']\partial_t^2\Psi-\triangle\Psi=0, \end{equation}
where $\triangle\equiv\partial_x^2+\partial_y^2+\partial_z^2$. Let us assume that the solutions are given by harmonic modes as
\begin{equation}\label{13}
\Psi_{n}(x,y,z,t)=N_{n}\exp{[i( k_yy+k_zz-\omega_{n} t)]}\sin{(n\pi x/L)}
\end{equation}
with the Dirichlet boundary conditions being satisfied on the conducting plates. Substituting equation (\ref{13}) into (\ref{12}), we find that the eigenfrequencies $\omega_{n}$ are given by the expression
\begin{equation}\label{14}
\omega_{n}=[1+\Phi_0'(b+1)]\left(k_y^2+k_z^2+\frac{n^2\pi^2}{L^2}\right)^{1/2}
\end{equation}

We proceed by calculating the normalization constant $N_n$, in order to find the vacuum expectation value of the $00$ component of the energy-momentum tensor. To do this, we will consider the norm  of the scalar functions $\Psi_n$ (which obey the usual orthonormality conditions), defined on the spacelike Cauchy hypersurface, $\Sigma$, as \cite{birrel},
\begin{equation}\label{15}
\|\Psi_n\|=-i\int_{\Sigma}\sqrt{-g_{\Sigma}}[\partial_t(\Psi_n)\Psi_n^{*}-\partial_t(\Psi_n^{*})\Psi_n]n^{t}d\Sigma
\end{equation}
where $g_{\Sigma}$ is the determinant of the metric induced on the hypersurface, $g_{ik}$, with $i,k = 1,2,3$ and $d\Sigma=dxdydz$ is its volume element. Following the procedure adopted by Sorge \cite{sorge}, we define the timelike future-directed unit vector $n^t$ as $n^t=(-g)^{1/4}(1,0,0,0)=[1+(1/2)\Phi_0'(b-3)](1,0,0,0)$. Thus, we find that the square of normalization constant is

\begin{equation}\label{16}
N^2_{n}=\frac{1}{(2\pi)^2\left(1+\frac{b-9}{2}\Phi_0'\right)L\omega_{n}}.
\end{equation}
The local vacuum expectation value of the scalar field energy density $T_{tt}$ is given by

\begin{equation}\label{17}
\epsilon_{vac}=<0|n^tn^tT_{tt}(\Psi_{n})|0>=\int d^2k_{||}\sum_nn^tn^tT_{tt}(\Psi_n)
\end{equation}
where $d^2k_{||}=dk_ydk_z$ and
\begin{equation}\label{18}
T_{tt}(\Psi_n)=\frac{1}{2}\left(\partial_t\Psi_n\partial_t\Psi^{*}_n-g_{tt}g^{ik}\partial_i\Psi_n\partial_k\Psi^{*}_n\right).
\end{equation}
The vacuum energy density $\overline{\epsilon}_{vac}$ is obtained by averaging $\epsilon_{vac}$ in the spatial
region (cavity) of the Casimir aparatus, whose expression is
\begin{equation}\label{19}
\overline{\epsilon}_{vac}=\frac{1}{V_p}\int_{\Sigma}\sqrt{-g_{\Sigma}}\epsilon_{vac}d\Sigma
\end{equation}
where $V_p=\int_{\Sigma}\sqrt{-g_{\Sigma}}d\Sigma$ is the proper volume of the cavity.
Putting (\ref{18}) into (\ref{19}) and solving the spatial integrals, we find
\begin{eqnarray}\label{20}
\overline{\epsilon}_{vac}=\left(1+\frac{3b+5}{2}\Phi_0'\right)\int\frac{d^2k_{||}}{2(2\pi)^2L}\sum_n\left[k^2_{||}+\left(\frac{n\pi}{L}\right)^2\right].\nonumber\\
\end{eqnarray}
The integral in the rhs of (\ref{20}) is just the vacuum energy density of the massless scalar field in Minkowski space,
which is infinity. The Casimir energy density $\overline{\epsilon}^{ren}_{vac}$ is found by using an adequate
renormalization technique, as the Abel-Plana subtraction formula \cite{bordag}, for example. Applying this renormalization process in equation (\ref{20}), we get
 \begin{equation}\label{21}
\overline{\epsilon}^{ren}_{vac}=-\left(1+\frac{3b+5}{2}\Phi_0'\right)\frac{\pi^2}{1440L^4}.
\end{equation}

We need to express this energy as a function of the proper length $L_p=\sqrt{g_{xx}}L\simeq(1-\Phi_0')L$.
Thus, in terms of the proper length $L_p$, the Casimir energy given by equation (\ref{21}) can be written as
\begin{eqnarray}\label{22}
\overline{\epsilon}^{ren}_{vac}&=&-\left(1+\frac{3b-3}{2}\Phi_0'\right)\frac{\pi^2}{1440L_p^4}\nonumber\\
&=&-\left(1+\frac{9M^2}{8\omega R^4}\right)\frac{\pi^2}{1440L_p^4}.
\end{eqnarray}
It is worth noticing that when $\omega\rightarrow\infty$, we recover the result obtained by \cite{sorge}, in the first order of approximation in $(M/R)$. In that case, the Casimir energy is the same one of the flat spacetime, i.e., the constant gravitational field does not affect the renormalized vacuum energy. Here, we have shown that, even in that order of approximation, the Casimir energy is modified due to the effects of Ho\v{r}ava-Lifshitz gravity.  Note that equation (\ref{22}) tells us that there is a reduction in this energy, which means that the force between the plates is intensified, which is according to the result found by Ferrari \textit{et al.} \cite{ferrari}, obtained using the perturbative approach.

\section{Thermal corrections}

Let us investigate now the role of the temperature on the Casimir energy by calculating its thermal corrections due to a thermal bath at absolute temperature $T$. For this task, we need to find the temperature-dependent renormalized Helmholtz free energy $\Delta_T\mathcal{F}_0^{ren}$, which is given by \cite{svaiter,bordag}
\begin{equation}\label{23}
\Delta_T\mathcal{F}_0^{ren}=2Ak_BT\sum_{n=0}^{\infty}\int\int\frac{d^2\textbf{k}_{||}}{(2\pi)^3}\log{\left[1-e^{-\widetilde{\beta}\left(\frac{n^2\pi^2}{L^2}+k^2_{||}\right)^{1/2}}\right]}-f_{bb}(T),
\end{equation}
where $A$ is the area of the plates, $k_B$ is the Boltzmann constant and $\widetilde{\beta}=[1+\Phi_0'(b+1)]/k_BT$, in accordance with equation (\ref{14}). The second term in the rhs of equation (\ref{22}) is the free energy density of black-body radiation, i.e., the non-confined part of the free energy density, expressed by
\begin{equation}\label{24}
f_{bb}(T)=k_BT\int\int\int\frac{d^3\textbf{k}}{(2\pi)^3}\log{\left[1-e^{-\widetilde{\beta}\left(k_x^2+k^2_{||}\right)^{1/2}}\right]},
\end{equation}
which can be easily found, in the approximations used here, as being given by
\begin{equation}\label{25}
f_{bb}(T)=-\frac{\left(1+\frac{6M}{R}-\frac{15M^2}{4\omega R^4}\right)\pi^2(k_BT)^4}{90}.
\end{equation}
We remark that, in absence of gravity, equation (\ref{25}) becomes the usual density of black body energy.

By using the Taylor's expansion for the logarithm, equation (\ref{23}) becomes
\begin{equation} \label{26}
\Delta_T\mathcal{F}_0^{ren}=\frac{Ak_BT}{\pi}\sum_{n,m=1}^\infty\frac{1}{m}\int_0^\infty dk_{||} k_{||} e^{-\widetilde{\beta}m\left(\frac{n^2\pi^2}{L^2}+k_{||}^2\right)^{1/2}}-f_{bb}(T),
\end{equation}
which, under an adequate change of variables, can be written as
\begin{equation}\label{27}
\Delta_T\mathcal{F}_0^{ren}=-\frac{Ak_BT\pi}{L^2}\sum_{n,m=1}^\infty\frac{n^2}{m}\int_1^\infty dx x e^{-\widetilde{\beta}mnx/L}-f_{bb}(T).
\end{equation}
Expressing the integral in terms of the modified Bessel's function of second kind \cite{grad}, we get
\begin{equation}\label{28}
\Delta_T\mathcal{F}_0^{ren}=-\frac{Ak_BT}{L^2}\sqrt{\frac{L}{\widetilde{\beta}}}\sum_{n,m=1}^\infty\left(\frac{n}{m}\right)^{3/2}K_{3/2}\left(\frac{\widetilde{\beta}mn\pi}{L}\right)-f_{bb}(T).
\end{equation}
Therefore, the total renormalized free energy is
\begin{eqnarray}\label{29}
\mathcal{F}_0^{ren}&=&AL_p\overline{\epsilon}^{ren}_{vac}+\Delta_T\mathcal{F}_0^{ren}\nonumber\\
&=&-F_1\frac{A\pi^2}{1440L_p^3}-F_2A(k_BT)^{3/2}L_p^{-3/2}\sum_{n,m=1}^\infty\left(\frac{n}{m}\right)^{3/2}K_{3/2}\left(\frac{\widetilde{\beta}mn\pi}{L}\right)+\nonumber\\
&+&F_3\frac{AL_p\pi^2(k_BT)^4}{90},\nonumber\\
\end{eqnarray}
where the factors related with the gravitational field are $F_1=\left(1+\frac{9M^2}{8\omega R^4}\right)$, $F_2=\left(1+\frac{3M}{2R}-\frac{3M^2}{8\omega R^4}\right)$, $F_3=\left(1+\frac{6M}{R}-\frac{15M^2}{4\omega R^4}\right)$ and $L_p$ is the proper distance between the plates, whose expression was obtained in the previous section.

\subsection{Internal energy - low and high temperature limits}
The next step is to calculate the renormalized internal energy, given by
\begin{equation}\label{30}
U_0^{ren}(T)=-T^2\frac{\partial{(\mathcal{F}^{ren}_0/T)}}{\partial T}.
\end{equation}
In order to obtain this physical quantity, we will take both low and high temperature limits of the total free energy (\ref{29}). For the first limit, we will consider the dominant term of the asymptotic expansion of the modified Bessel's function for great arguments \cite{grad}, and thus we get
\begin{equation}\label{31}
U_0^{ren}(T)\thickapprox-F_1\frac{A\pi^2}{1440L_p^3}+\frac{A\left[F_4\pi L_p^{-1}(k_BT)+F_5(k_BT)^2\right]}{\sqrt{2}L_p}e^{-\widetilde{\beta}\pi/L}-F_3\frac{AL_p\pi^2(k_BT)^4}{30},
\end{equation}
where $F_4=\left(1+\frac{M}{R}+\frac{M^2}{2\omega R^4}\right)$ and $F_5=\left(1+\frac{2M}{R}-\frac{M^2}{2\omega R^4}\right)$. We note that when $T\rightarrow0$, the renormalized internal energy coincides with the Casimir one at $T=0$.

The high temperature limit is obtained from the asymptotic expansion of the modified Bessel's function for small arguments \cite{grad}. Thus we have
\begin{equation}\label{32}
U_0^{ren}(T)\thickapprox-F_1\frac{A\pi^2}{1440L_p^3}-F_6\frac{A\zeta(3)(k_BT)^3}{\pi}-F_3\frac{AL_p\pi^2(k_BT)^4}{30},
\end{equation}
where $F_6=\left(1+\frac{3M}{2R}-\frac{15M^2}{8\omega R^4}\right)$. It's interesting to note that the second term in equation (\ref{31}) does not contribute to the thermal Casimir force $-\frac{\partial U_0^{ren}}{\partial L_p}$. The dominant term arises from the black body radiation.

\section{Concluding remarks}
 We have calculated the Casimir energy of a massless scalar field in two nearby, parallel, uncharged conductor plates placed tangentially to the surface of a sphere with mass $M$ and radius $R$. The analysis taken into account the effects due to a static, spherically symmetric and asymptotically flat background gravity \textit{\`a la} Ho\v{r}ava-Lifshitz described by Kehagias-Sfetsos solution. For this analysis, we considered the approximation in first order of $1/R$ (weak field) and $1/\omega$ (IR scale), i.e., emphasizing the lower order of both Schwarzschild and Kehagias-Sfetsos effects. The dynamics of the scalar field, in this approximation in which $\omega\gg1$ and $z\thickapprox1$, is the usual, locally obeying the Lorentz symmetry, therefore.

We showed that the Minkowski Casimir energy is not modified in the first order of approximation of $(1/R)$, consistent with a constant Newtonian gravitational potential, which reproduces the results of \cite{sorge}, but it is altered in first order of $1/\omega$, being subtracted by an amount $(9M^2/8\omega R^4)\overline{\epsilon}^{ren,flat}_{vac}$. This means that the equivalence principle is not obeyed in such scenario, as expected in the context of Ho\v{r}ava-Lifshitz gravity, since that eq. (22) expresses a renormalizable quantum vacuum energy that is under influence of a constant gravitational potential (strictly local, therefore), which is different of the Casimir energy obtained in flat spacetime, due to the leading term of KS solution, and the equality between these energies is required by the equivalence principle. If the free parameter $\omega$ of the mentioned term goes to infinite, the equality is established, and then one recovers the Einstein's gravity with all its principles and symmetries. We also call attention to the fact that the attractive Casimir force is amplified with respect to the Minkowski one, in accordance with the result that comes from the perturbative approach found in \cite{ferrari}.

 Inserting in the correction found by us the factor $G^2/c^4$, taking the parameters of Earth and the current uncertainty of measurements involving the Casimir effect, which is of order of $10^{-2}$ \cite{mohideen1,mohideen2}, we can estimate a lower bound to $\omega$, which is such that $\omega \geqslant10^{-34}cm^{-2}$. We need to confront this limit with the constraint obtained from the classical tests of general relativity (perihelion precession of Mercury, deflection of light by the Sun and the radar echo delay) in solar system, in which $\omega\geqslant10^{-27}cm^{-2}$ \cite{harko}. This comparison means that the mentioned measurements made in solar system are much more precise than the terrestrial Casimir ones, and that we must improve in least $7$ orders of magnitude the precision of the latter (making it compatible with other tests of Quantum Electrodynamics \cite{peskin}) if we want to have a real chance of finding signals of Horava-Lifshitz gravity in this kind of experiment. We also remark that the growing of Casimir force between the plates due to the Ho\v{r}ava-Lifshitz gravity term can be seen as a local effect related to the negative cosmological constant derived from this theory \cite{corrado}. Thus, we consider our approach consistent, do not finding any possibility of change in the signal of the Casimir force neither residual corrections for it at IR limit, as found in \cite{ferrari}.

We have also examined the role played by the temperature in the Casimir effect, calculating the renormalized free energy and, consequently, the internal energy of the system, in the low and high temperature limits. We point out that the obtained results contain naturally the thermal extension of the paper \cite{sorge}, when the KS parameter $\omega\rightarrow\infty$, and it's interesting to note that, despite the gravitational field does not manifest itself in the Casimir effect when we consider only terms $\mathcal{O}(M/R)$ at $T=0$, it shows up in the thermal corrections.
\section*{Acknowledgements}
C.R. Muniz would like to thank Universidade Federal da Para\'iba (UFPB) for the kind welcome and CNPq for the Postdoctoral Fellowship. V.B.Bezerra would like to thank CNPq for partial financial support. M.S. Cunha acknowledge warm hospitality at the UFPB.

\end{document}